# Hydrothermal Alteration at the Panorama Formation, North Pole Dome, Pilbara Craton, Western Australia

Adrian J. Brown[*,1], Thomas J. Cudahy[2,3] and Malcolm R. Walter[3]

[1] SETI Institute, 515 N. Whisman Rd, Mountain View, CA 94043, USA
[2] CSIRO Exploration and Mining, ARRC Centre, Kensington, WA 6102, Australia
[3] Australian Centre For Astrobiology, Macquarie University, NSW 2109, Australia

**ABSTRACT**

An airborne hyperspectral remote sensing dataset was obtained of the North Pole Dome region of the Pilbara Craton in October 2002. It has been analyzed for indications of hydrothermal minerals. Here we report on the identification and mapping of hydrothermal minerals in the 3.459 Ga Panorama Formation and surrounding strata. The spatial distribution of a pattern of subvertical pyrophyllite rich veins connected to a pyrophyllite rich palaeohorizontal layer is interpreted to represent the base of an acid-sulfate epithermal system that is unconformably overlain by the stromatolitic 3.42Ga Strelley Pool Chert.

**KEYWORDS**

[*] corresponding author, email: abrown@seti.org

# INTRODUCTION

Hyperspectral remote sensing has been used in the past to successfully survey Archean regions of Western Australia (Cudahy, et al., 2000). As part of a combined Australian Centre for Astrobiology-CSIRO project to map the North Pole Dome in the Pilbara region, a remote sensing dataset was collected in October 2002 and has been analyzed in the years hence. The dataset was obtained using the reflectance spectrometer called HyMap (Cocks, et al., 1998), owned and operated by HyVista Corporation. This instrument measures reflected light from 0.4 to 2.5 microns, and is particularly suited to mapping hydrothermal minerals due to 2.2-2.35 micron vibrational absorptions of the $OH^-$ anion in alteration minerals (Hunt, 1979).

# GEOLOGICAL SETTING

The 3.5Ga Pilbara Craton covers 60,000km$^2$ and provides some excellent exposure of low metamorphic grade, relatively undeformed rocks that span much of the Archean (ca. 3.5-2.7Ga). The Pilbara Craton is divided into three granite-greenstone terranes with distinct geological histories, separated by intervening clastic sedimentary basins (Van Kranendonk, et al., 2002). The Pilbara is an important region for investigating early Earth processes, but its state of preservation has also been valuable to investigate processes relevant to all time

*Adrian J. Brown*

periods, for example the study of VHMS systems at the Strelley Granite (Cudahy, et al. 2000).

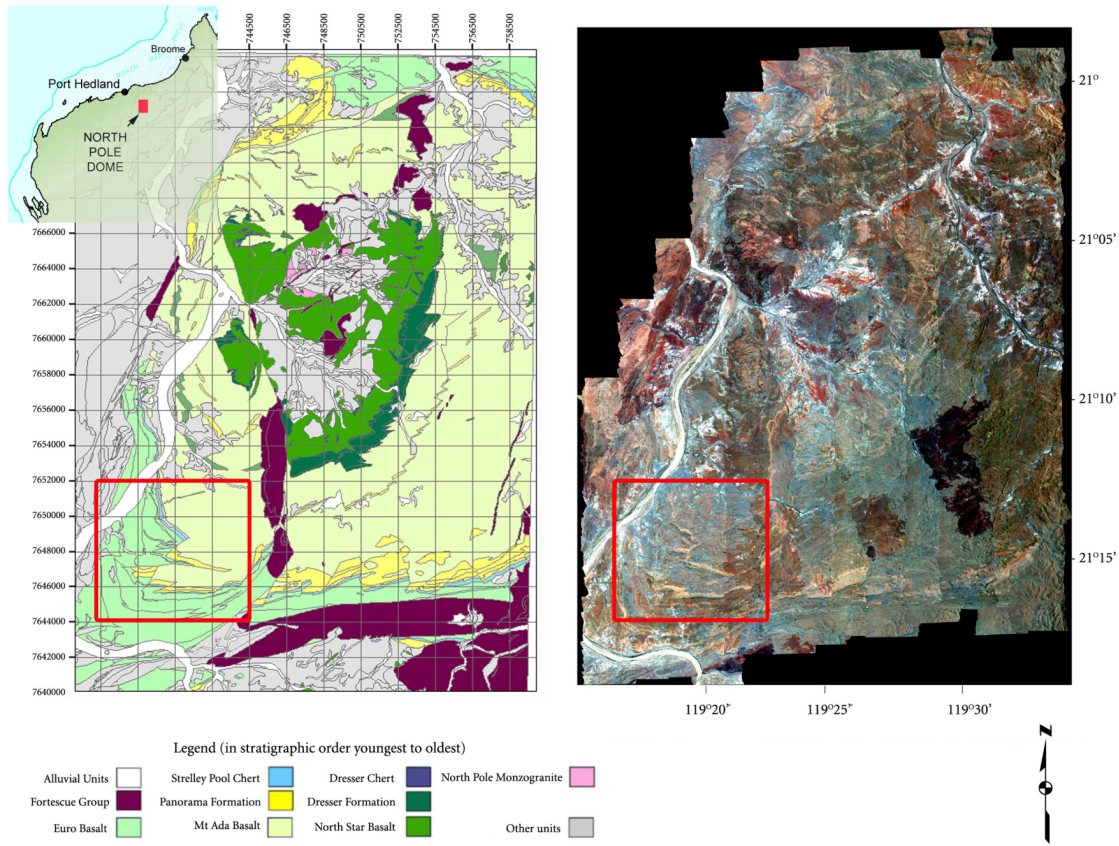

Figure 1 – North Pole Dome with red box outlining the best preserved regions of the Panorama Formation. Grid lines are Australian Map Grid (AMG84 or UTM Zone 50). Each grid box is 2km across. Mineral maps in Figure 4 are sourced from the region within the red box.

The North Pole Dome (NPD) is located in the center of the East Pilbara Granite Greenstone Terrain (Van Kranendonk, 2000). The NPD is a structural dome of bedded, dominantly mafic volcanic rocks of the Warrawoona and Kelly Groups that dip gently away from the North Pole Monzogranite exposed in the core of the Dome (Figure 1; Van Kranendonk, et al., 2002). Average dips vary from 50-60° in the inner Dresser Formation to around 30-40° in the outer Panorama Formation (Van Kranendonk, 2000). The North Pole Monzogranite has been estimated to extend ~6km below the surface by gravity surveys (Blewett, et al., 2004). The central monzogranite has been interpreted as a syn-volcanic laccolith to Panorama Formation volcanic rocks at the top of the Warrawoona Group (Blewett, et al., 2004, Van Kranendonk, 1999). Minor outcrops of felsic volcanic rocks are interbedded with the greenstones, and these are capped by cherts that indicate hiatuses in volcanism (Barley 1993, Van Kranendonk, 2000). An overall arc-related model for hydrothermal activity is favored by Barley (1993), whereas Van Kranendonk has recently supported a mantle-plume driven model for hydrothermal activity at the North Pole Dome (Van Kranendonk, 2006).

A stratigraphic column for the study area (Figure 2) shows the age and composition of the main stratigraphic units of the Kelly and Warrawoona Groups. Formation codes are given in parentheses. The region is regarded as an important witness to Earth's earliest biosphere since stromatolite and possible microfossil occurrences have been documented at three distinct stratigraphic levels at the NPD – within the Dresser Formation, the Mount Ada Basalt and the Strelley Pool Chert (Dunlop, et al., 1978, Walter, et al. 1980, Awramik, et al.,1984, Ueno, et al. 2001, Buick, 1990, Van Kranendonk, 2006, Ueno, et al. 2004).

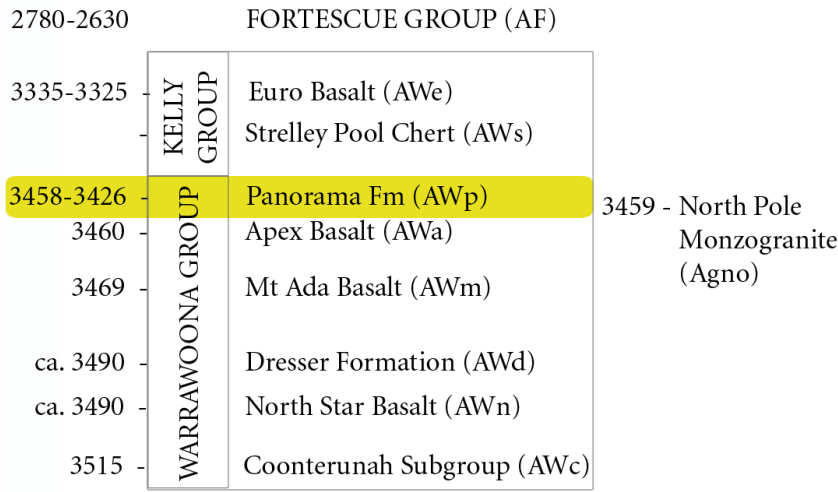

Figure 2 – Stratigraphic column of the geological units within the study region.

**<u>*Warrawoona Group - Panorama Formation (AWp)*</u>**. The Panorama Formation consists of felsic volcanic rocks that outcrop as a prominent ridge line around the outer part of the NPD. It is up to 1km thick and has been extensively silicified (Cullers, et al., 1993). It has been directly dated (using U-Pb in zircon) as between 3.458 and 3.426 Ga (Van Kranendonk, et al., 2002). The rhyolitic Panorama Formation was erupted after the Mt Ada ultramafic-mafic sequence and Apex Basalt. The variation in style from ultramafic through mafic to felsic volcanics at the NPD was possibly driven by a mantle plume that also drove hydrothermal activity during this period (Van Kranendonk, 2005). Remnants of the felsic volcanic rocks of the Panorama Formation are distributed 7-16km from the center of the North Pole Dome. Some of the most accessible Panorama Formation elements lie in an 8x8km region on the southwest corner of the Dome, and these make up the study region of this paper.

***Kelly Group (AK) - Strelley Pool Chert (AKs)***. The Strelley Pool Chert (SPC) is a marker unit that crops out in greenstone belts across the East Pilbara (Van Kranendonk, 2000). It chiefly consists of a laminated grey to black and white chert that represents silicified carbonate, commonly with coniform stromatolites. It has not been directly dated, but was deposited between 3.426 and 3.335 Ga, the ages of underlying and overlying felsic volcanic rocks (Van Kranendonk, 2006). In the study region, it unconformably overlies the Panorama Formation.

Relationship of Panorama Formation to central NP Monzogranite. Palaeocurrent evidence from crossbedded arenites in the Panorama Formation suggest sediment dispersal from a felsic source close to the modern day center of the NPD (DiMarco, 1989). It has recently been suggested that an eruptive vent of this felsic volcanic source is preserved in the north western part of the NPD (Van Kranendonk, 1999). Van Kranendonk (1999) opined that the NP Monzogranite fed the 'Panorama Volcano' based on similar ages (Thorpe, et al., 1992) and the presence of radially oriented dykes leading from the granite to the level of the Panorama Formation (Van Kranendonk, 2000).

Hydrothermal activity associated with the Panorama Formation. Following deposition of the Mt Ada Basalt, and during or soon after the deposition of the Panorama Formation, an intense period of hydrothermal alteration occurred at the palaeosurface (probably in a shallow submarine setting). A highly schistose, intensely altered, golden brown pyrophyllite-rich horizon represents the alteration event (Van Kranendonk and Pirajno, 2004) that extends around the edge of the NPD, in the contact between the Mount Ada Basalt and the Panorama Formation, and in and adjacent to granite dykes radiating out from the North Pole Monzogranite.

The hydrothermal event and subsequent deformation that gave rise to the pyrophyllite schists in the Mt Ada Basalt and Panorama Formation predates deposition of the Kelly Group, as the former are affected by a penetrative schistosity and high temperature alteration that is absent from the unconformably overlying rocks of the Kelly Formation (Van Kranendonk and Pirajno, 2004, Brown, et al., 2004). The top of the alteration system, as well as structurally higher parts of the alteration system may have been eroded away before emplacement of the overlying Kelly Group – in the form of the downcutting SPC.

## REMOTE SENSING OF HYDROTHERMAL MINERALS AND ALTERATION ZONES

<u>Detection of vibrations due to $OH^-$</u>. Electromagnetic radiation that intercepts the surface of a mineral bearing the $OH^-$ cation (often called 'alteration' or hydrothermal minerals) will be absorbed at certain wavelengths due to sympathetic vibrations within the crystal lattice (Hunt, 1977). The electronic neighborhood of the $OH^-$ cation affects the energy of the vibration, and hence the central absorption wavelength, of the reflected light. In this manner, Al-OH bonds in muscovite, for example, can be unequivocally distinguished from Mg-OH bonds in chlorites or amphiboles. Photons with wavelengths between 2.0-2.5 microns have similar energies to combinations of fundamental OH$^-$ vibrations, and thus are suitable for detecting hydrothermal minerals.

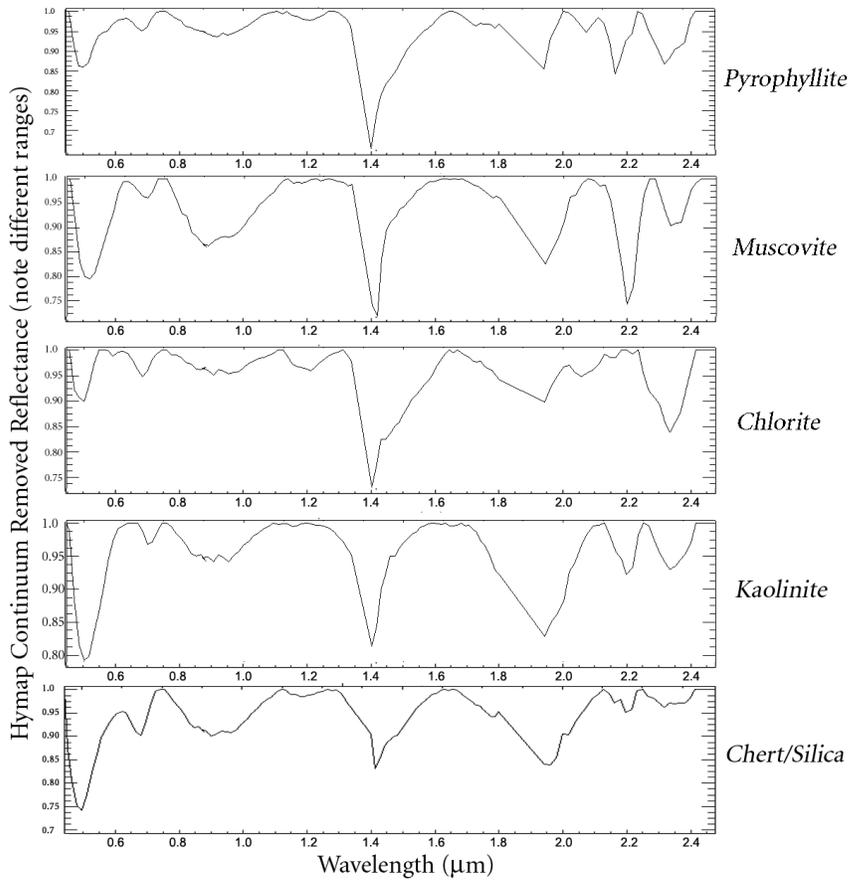

Figure 3 – Representative HyMap continuum removed reflectance spectra for OH- bearing minerals and silica taken from the hyperspectral dataset of the study region.

Figure 3 shows an example of Hymap spectra of five different minerals. The 2.0-2.5 micron region of each of these spectra is unique and diagnostic of the first four minerals. Silica, the last mineral, is identified by an absence of strong bands in this region. Field mapping has shown this to be a successful strategy in this highly silicified granite-greenstone region. This strategy should be used with caution in other regions, where silicification and alteration minerals are rare, since it does not discriminate between silica and unaltered feldspar or pyroxene, for example. Each of the spectra in Figure 3 has been 'continuum removed' in order to highlight the absorption bands (Clark, et al., 1987).

In order to map the diagnostic absorption bands, a computer algorithm has been developed to fit Gaussian-type curves to the absorption bands in the 2.0-2.5 micron region. This process is discussed in more detail in companion papers (Brown, 2005, Brown 2006a). We used the IDL programming language and the ENVI environment (www.rsi.com) to produce mineral maps of hydrothermal minerals with diagnostic absorption bands. Each hydrothermal mineral has been chosen as a representative of an alteration zone type on the basis that it is the spectrally dominant contributor to the mineral assemblage (Thompson and Thompson, 1996). The alteration zones and mineral assignments are given in Table 1.

| Mineral | Alteration Zone | | | |
|---|---|---|---|---|
| | Phyllic | Advanced Argillic | Propylitic | Silicic |
| Muscovite | ■ | | | |
| Pyrophyllite | | ■ | | |
| Chlorite | | | ■ | |
| Silica/Chert | | | | ■ |

Table 1 – Alteration zones and associated minerals in the Panorama region.

<u>Al-OH absorption band wavelength</u>. The central wavelength of vibrational absorption bands is very sensitive to the local chemical bonding environment. The central wavelength of the 2.2 micron Al-OH vibration absorption band can be used as a proxy for Al-content in white micas such as muscovite (Duke, 1994). It has been used to track fluid pathways in hydrothermally altered rocks (van Ruitenbeek, et al., 2005) and as a measure of metamorphic grade in muscovitic schists (Longhi, et al., 2000). Long wavelength white micas have lower Al content (higher Si content) and short wavelength white micas have higher Al content.

## FIELD SAMPLING

Three field trips to the Panorama Formation region from 2003-2005 yielded over 85 samples which have been used to ground-truth hyperspectral mineral maps presented herein (Figure 4). All major units (including the Euro Basalt, Mt Ada Basalt, Panorama Formation and Al-rich and poor muscovite veins) in the study area were sampled, particularly those units which displayed spectral heterogeneity. Sample locations are shown on the geological map in Figure 4. The results of XRD, XRF, IR and EMP analyses of these samples are discussed elsewhere (Brown 2006b). Hyperspectral mineral identifications were confirmed mostly using petrographic slides, in particular the identification of the endmember minerals silica, pyrophyllite, muscovite and chlorite. Muscovite was distinguished from pyrophyllite using XRD of small amounts of powdered samples. Al-rich and Al-poor muscovite were distinguished using EMP analysis.

*Adrian J. Brown*

We used a PIMA SWIR hand-held spectrometer (www.hyvista.com) to obtain laboratory IR spectra of all our samples (Brown, et al., 2004). These laboratory spectra were used to compare the spectral signature of samples to the signatures of hydrothermal minerals in the airborne dataset.

## RESULTS

Mineral maps are presented in Figure 4, along with a 'hydrothermal facies' map where individual mineral maps are combined. A geological map of the region showing sample locations is also provided.

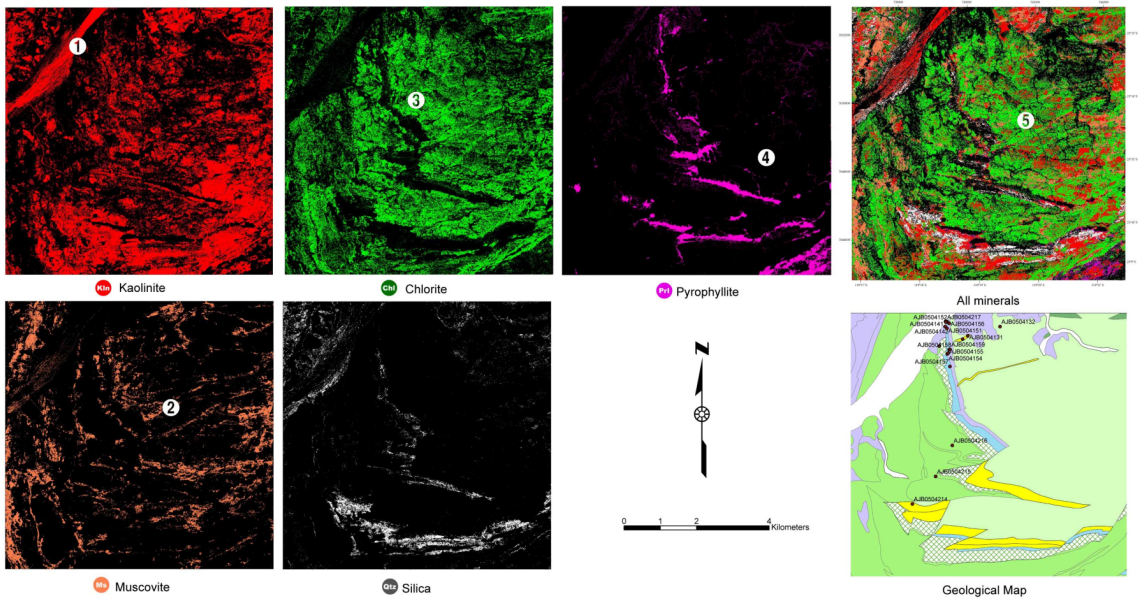

Figure 4 - Mineral maps of the study region, showing distributions of the indicated minerals. See Figure 1 for coverage. The upper right image is a complete hydrothermal zone map, formed by combining the other five other maps together. Some points are obscured – pyrophyllite was laid down first, then silica, then chlorite, then kaolinite, then muscovite. This is an alteration class map – pink is the phyllic zone, white silicic, red argillic, green propylitic, and purple advanced argillic. Part of the geological map from Figure 1 is provided (Van Kranendonk, 2000) of the region showing sample locations – this shows a silica rich region of the SPC in hatched green.

1. <u>Intense kaolinite in the Shaw River</u>. The dry, dusty and sandy river bed of the Shaw River is easily discriminated by its consistent kaolinite signature. Using a kaolinite index linked to the sharpness of the 2.16-2.2 micron doublet of kaolinite (Brown, 2006) suggests that kaolinite in the Shaw River is detrital, and probably due to weathering of felsic Al rich material from nearby granites and phyllosilicate bearing sandstones.

2. <u>Arcuate regions of phyllic alteration</u>. Phyllic alteration (defined here by the presence of muscovite; Table 1) has two styles in this region: 1.) long, thin arcs of ridge-forming "bedding parallel" phyllic alteration follow the general outline of ridges, and 2.) long thin veins leading up to the arcuate ridges have zones of phyllic alteration. There is a large amount of phyllic alteration on the extreme west of the study area, due to sandstone units of the Gorge Creek Supergroup, outside the NPD (Figure 4).

3. <u>Pervasive propylitic alteration beneath and above the Panorama Formation</u>. Propylitically altered, chlorite-bearing pillow basalts of the Mt Ada Basalt and overlying Euro Basalt make up most of the study area. The Euro Basalt has experienced only low strain, low temperature (propylitic) hydrothermal alteration and very low metamorphic grade (Van Kranendonk, 2000).

4. <u>Advanced Argillic alteration</u>. The determining feature of the Panorama hydrothermal event is the advanced argillic layer defined by the presence of pyrophyllite. Pyrophyllite is largely developed in at least two rock types – the felsic Panorama Formation and the top parts of the underlying mafic Mt Ada Basalt. Pyrophyllite bearing units outcrop beneath, and do not extend into, the overlying Euro Basalt, indicating that the timing of the pyrophyllite development was prior to the Euro Basalt deposition. The pyrophyllite bearing units are characterized by highly schistose, golden brown weathering habit. This alteration style was previously interpreted as representing medium temperature (300 °C), alteration with a high strain event during, or soon after, emplacement (Van Kranendonk and Pirajno, 2004).

5. <u>Juxtaposition of hydrothermal zones</u>. Pyrophyllite rich alteration is everywhere developed below the SPC. In addition, it should be noted that the phyllic alteration at the base of veins leading up to the SPC grades from muscovite to pyrophyllite rich, as discussed below.

*Muscovite Al-OH Wavelength Map*

As discussed earlier, white mica Al content can be mapped by tracking the central wavelength of the Al-OH feature near 2.2µm (Duke, 1994). Using computer algorithms discussed elsewhere (Brown, 2006a), we have generated a muscovite Al-OH wavelength map and overlain the pyrophyllite mineral map (Figure 5).

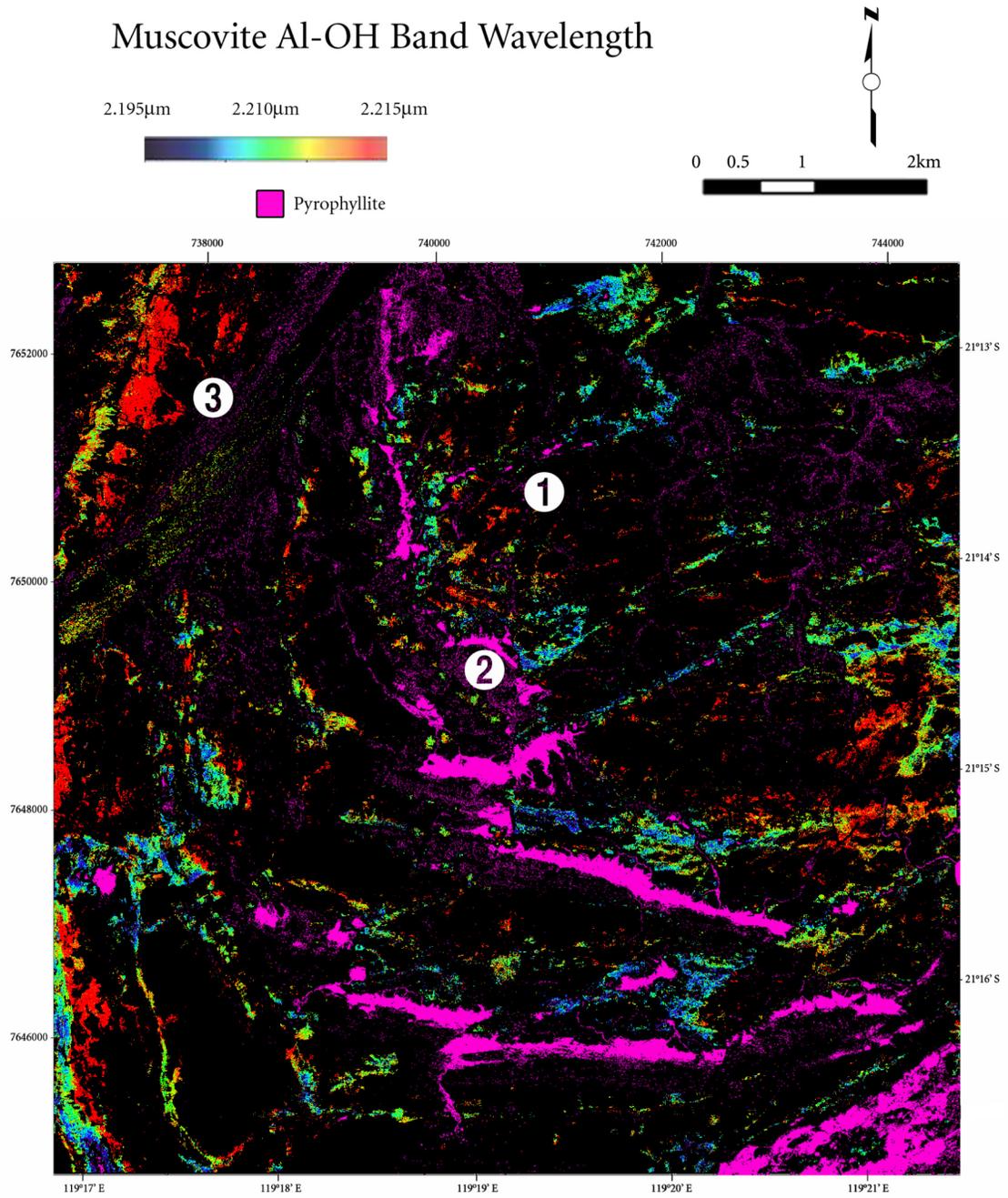

Figure 5 – Muscovite Al-OH Wavelength map for the Panorama region. Long wavelength (Al-poor) muscovite is in red and short wavelength (Al-rich) muscovite is in blue. Pyrophyllite is in pink. The area covered is identical to the maps in Figure 4.

1. <u>Variations in Al content of white micas beneath pyrophyllite in Panorama Formation</u>. Around point ι on Figure 5, red colors indicate the presence of long wavelength (Al-poor) white mica just beneath the Panorama Formation. Moving east, linear muscovite veins are represented by progressively bluer colors, indicating a shift to shorter wavelength (Al-rich) muscovite. This pattern also repeats on the extreme east of the map, where red Al-poor veins grade to green and blue Al-rich muscovite. This pattern is not entirely uniform, there are, for example, blue patches of Al-rich muscovite beneath the Panorama Formation (one Al-rich vein terminates near φ) that display fairly constant blue color.

2. <u>Pyrophyllite-rich sub-horizontal zones of the Panorama Formation</u>. Overall, the muscovite veins described above tend to start Al-rich (blue) in the east, and become more Al-poor (red), and then merge into pyrophyllite sub-vertical veins which then connect with pyrophyllite-rich sub horizontal zones. In most regions immediately below the pyrophyllite, there is a thin sub-horizontal layer of Al-rich (blue) muscovite, for example at point φ. This thin horizontal blue layer somewhat hides the nature of the muscovite and pyrophyllite mixing in the veins.

3. <u>Gorge Creek Supergroup alteration</u>. On the extreme left edge of the image, running north-south, a fault zone divides the phyllic alteration of the Gorge Creek Supergroup on the west side from the Euro Basalt (bearing no muscovite) on the right. The younger rocks in the west display well developed phyllic alteration (almost 1km wide in places) with variations in white mica Al-content from base to top (see also western edge and south west corner of this image).

4. <u>Pyrophyllite indicating structural doubling of Panorama Formation</u>. In the bottom of Figure 5, the occurrence of pyrophyllite is showing the structural faulting and doubling of the same horizon. The patch of pyrophyllite in the south east corner represents a larger exposure of the Panorama Formation controlled by the north-south trending Antarctic Fault which cuts through the NPD (Van Kranendonk, 2000). As can be seen by comparing it to the Van Kranendonk's geological map in Figure 4, this horizon was not mapped by him – it has been interpreted previously as Euro Basalt. The structural repetition of the Panorama Formation in the south of the study area is also shown on the pyrophyllite mineral map (Figures 4 and 5).

## DISCUSSION

<u>Background propylitic alteration</u>. The mafic pillow rocks of the underlying Mt Ada Basalt have all been extensively altered to chlorite-carbonate, possibly via the reaction described by equation 1 in Table 2. The presence of hydrothermal chlorite unfortunately does not uniquely distinguish them amongst the chloritically altered basalts present throughout the Warrawoona Group and in the overlying Kelly Group (Figure 2, to the left of point $\kappa$).

Heat source for the Panorama Hydrothermal Event. Van Kranendonk and Pirajno (2004) suggested that the heat source that drove silicic hydrothermal alteration of the SPC and Panorama Formation was the thick komatiitic basalt of the Euro Basalt on top of the SPC. This scenario seems unlikely following this study, given the presence of thin (but spatially extensive), relatively unaltered and unfoliated erosional units wedged in various locations in the study area between the pyrophyllite unit and the SPC. The similar alteration style in muscovite bearing veins leading up to the pyrophyllite layer suggests that the phyllic and pyrophyllite alteration was generated by a subvolcanic heat source, specifically the North Pole Monzogranite. What is not clear is how far below the palaeosurface the pyrophyllite layer was developed and how much of the top part of the system (including possible sinters, vents and exhalative mineralization which have not been found to date) has been eroded away by the unconformity at the base of the Kelly Group.

Occurrence of pyrophyllite. Pyrophyllite is a rock-forming mineral that occurs in a limited range of metamorphic and hydrothermal conditions. The primary condition for its petrogenesis is high Al – more Al than can be accommodated by the relatively aluminous alteration minerals muscovite, chlorite, chloritoid and lawsonite (Evans and Guggenheim, 1988). It commonly occurs in rocks that are enriched (relatively) in Al through base leaching during hydrothermal alteration, and is precipitated in hydrothermal veins (Evans and Guggenheim, 1988). It appears in both these circumstances in the Panorama Formation.

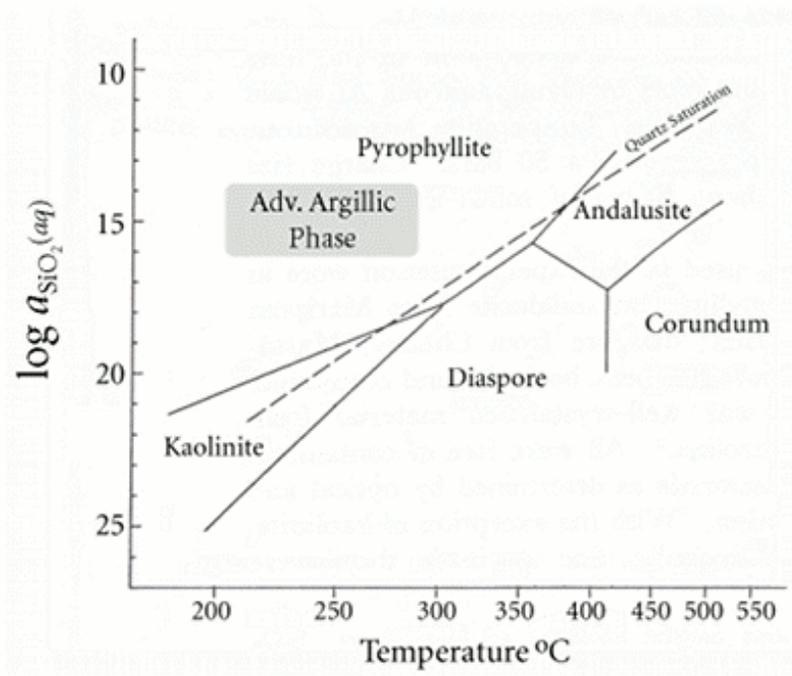

Figure 6 – Phase diagram of the Al$_2$O$_3$-SiO$_2$-H$_2$O system at 1-2kbar after (Hemley, et al., 1971).

The phase diagram for the Al$_2$O$_3$-SiO$_2$-H$_2$O system is shown in Figure 6 (Hemley, et al., 1980). This shows the control that high silica fluids have over whether pyrophyllite or kaolinite is precipitated from this system at low temperature. In fluids with high amounts of dissolved silica, pyrophyllite is the preferred phase.

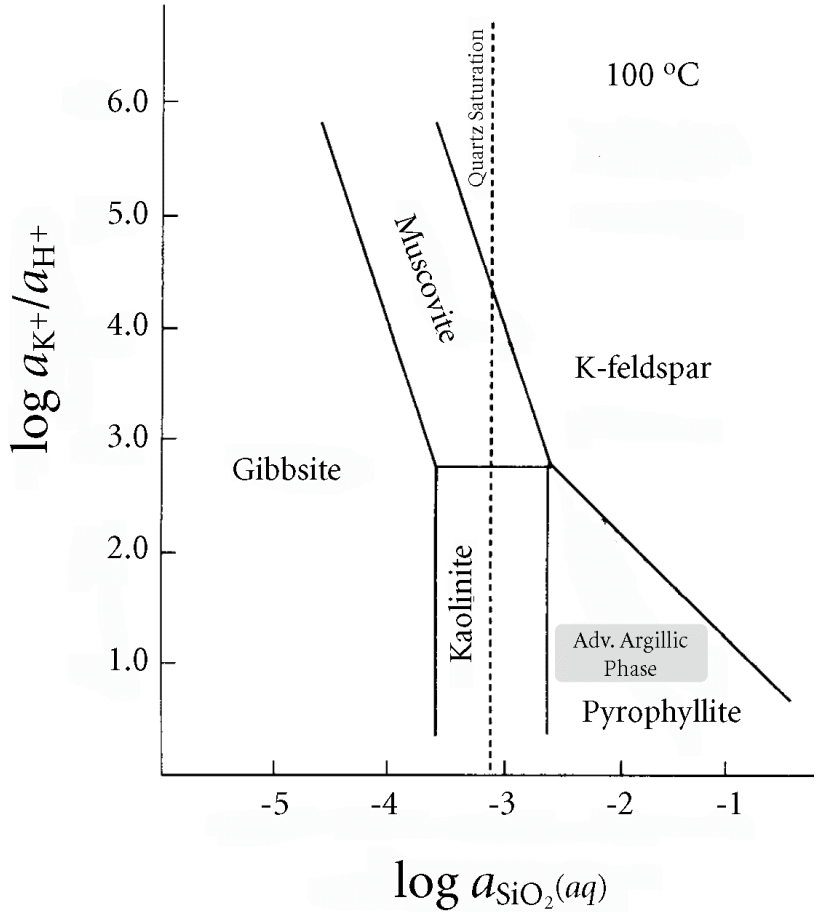

Figure 7 – Activity diagram of log $a_{K+}/a_{H+}$ versus silica content at 100° C. Generated using computer code SUPCRT92 after (Johnson, et al., 1992).

The activity diagram of activity of $K^+$/activity of $H^+$ versus silica content for the $Al_2O_3$-$SiO_2$-$H_2O$ system is shown in Figure 7 (Johnson, et al., 1992). This diagram was generated using the SUPCRT92 computer code (Johnson, et al., 1992). The diagram shows that in high silica and low pH conditions, pyrophyllite will be formed in preference to kaolinite.

Combining information from Figure 6 and 7, at near surface conditions (100°C and 1 atm) pyrophyllite is more stable than kaolinite (and muscovite) in waters supersaturated in silica (Hemley, et al., 1980). The occurrence of pyrophyllite at the top of sub-vertical muscovite veins may indicate a drop in temperature, a drop in oxygen fugacity, or an increase in silica content in the fluids or wall rock.

| Eqn # | Hydrothermal Zone | Chemical Reaction | Temperature (°C) | Conditions | Reference |
|---|---|---|---|---|---|
| 1 | Propylitic or Greenschist | $5Mg^{+2}+CaAl_2Si_2O_8+8H_2O+SiO_2 \Rightarrow Mg_5AlSi_3AlO_{10}(OH)_8+Ca^{+2}+8H^+$<br>*anorthite*     *chlorite* | 100-200 |  | (Hemley, et al., 1980) |
| 2 | Phyllic | $3\ KAlSi_3O_8 + H^+ \Leftrightarrow KAl_2(Si_3Al)O_{10}(OH)_2 + 6SiO_2 + K^+$<br>*k-fspar*          *muscovite*           *quartz* | 100-300 | pH >2 | (Montoya and Henley, 1975) |
| 3 | Advanced Argillic | $2KAl_2(Si_3Al)O_{10}(OH)_2 + 6SiO_2 + 2H^+ \Leftrightarrow 3Al_2Si_4O_{10}(OH)_2 + 2K^+$<br>*muscovite*                                *pyrophyllite* | 100 or 200-300 | pH ~ 2 | (Evans and Guggenheim, 1988) |

Table 2 – Probable hydrothermal alteration reactions, temperatures and conditions to produce observed mineral assemblages in the Panorama Formation region.

pH conditions of the hydrothermal system. One possible explanation for the observed mineral variation is that acidic $CO_2$-rich vapor was released as the fluids ascended and decompressed, attacking and altering the upper parts of the system to an advanced argillic stage whereas the lower parts remained at the phyllic stage (White and Hedenquist, 1990). Chemical equations 2 and 3 of Table 2 provide a possible pathway for the coexistence of muscovite and pyrophyllite in this system. In more acidic, silica-rich parts of the system, pyrophyllite is favored over muscovite. In less acidic conditions, muscovite predominates.

Redox conditions of the Panorama hydrothermal system. Recent research at the CSIRO Division of Exploration and Mining suggests pyrophyllite may be preferred to kaolinite or muscovite in reducing acidic conditions (J. Walshe, pers. comm.). It is possible that a reducing ocean may have been the source of hydrothermal fluids in the top, pyrophyllite rich part of the Panorama hydrothermal system (Nakamura and Kato, 2004).

Silica content of hydrothermal fluids. A final factor controlling the deposition of pyrophyllite and muscovite in the vein systems is the silica content of the fluids. An increase in silica at the paleosurface may have been driven by a high silica Archean ocean (Knauth and Lowe, 2003). Silica rich water drawn down into the top reaches of the hydrothermal system may have allowed the precipitation of pyrophyllite in preference to muscovite. The highly siliceous Panorama Formation may also have been a source of silica in the veins of this hydrothermal system.

Panorama hydrothermal event as an acid-sulfate system. An acid-sulfate type system is favored to describe the Panorama Hydrothermal Event primarily due to the intense advanced argillic alteration, driven by acidic, reducing fluids, which is consistent with the preserved mineral assemblages.

The muscovite-altering-to-pyrophyllite patterns in Figure 5 provide evidence for a change in fluid chemistry within the deep penetrating (1-4km) sub-vertical veins. The clear connection of pyrophyllite and muscovite in these veins, and the variations of these two minerals suggests a genetic link between the sub vertical and sub horizontal zones of the Panorama Formation, with alteration styles changing across-strike (up stratigraphy) due to temperature, acidity or redox conditions in a continuously connected hydrothermal system approaching the paleosurface.

It is possible the pyrophyllite alteration of the Panorama Event occurred in low temperature (100°C) conditions (Gradusov and Zotov, 1975). Temperature is not a well constrained parameter of acid-sulfate systems (Heald, et al., 1987). It is of note that salinities of acid-sulfate systems are commonly quite high (usually 5-24 % NaCl equivalent (Hayba, et al., 1985)). An important alteration mineral in acid-sulfate systems is alunite – this has not been found in this study, though it was reported by (Van Kranendonk and Pirajno, 2004) in the Strelley Pool Chert level. Identification of characteristic minerals such as alunite and jarosite in future studies would support the moniker of "acid-sulfate" for this hydrothermal system.

## PETROGENETIC MODEL

A three-stage model may best describe the genesis of the rocks within the study region for the development of main geological features in the study area is proposed:

1. <u>Advanced Argillic Phase</u>. Following shallow intrusion of the North Pole Dome Monzogranite, quartz porphyry veins were emplaced into the overlying crust and fed surface volcanic rocks of the Panorama Formation. Pertinent features of epithermal ore deposits formed under similar conditions are represented in the schematic diagram shown in Figure 8 (Bonham Jr. and Giles, 1983). The figure shows a subaerial system, however a shallow marine to occasional subaerial environment has been suggested by REE systematics and eruption patterns of the Panorama Volcano (Van Kranendonk and Hickman, 2000, Bolhar, et al., 2005).

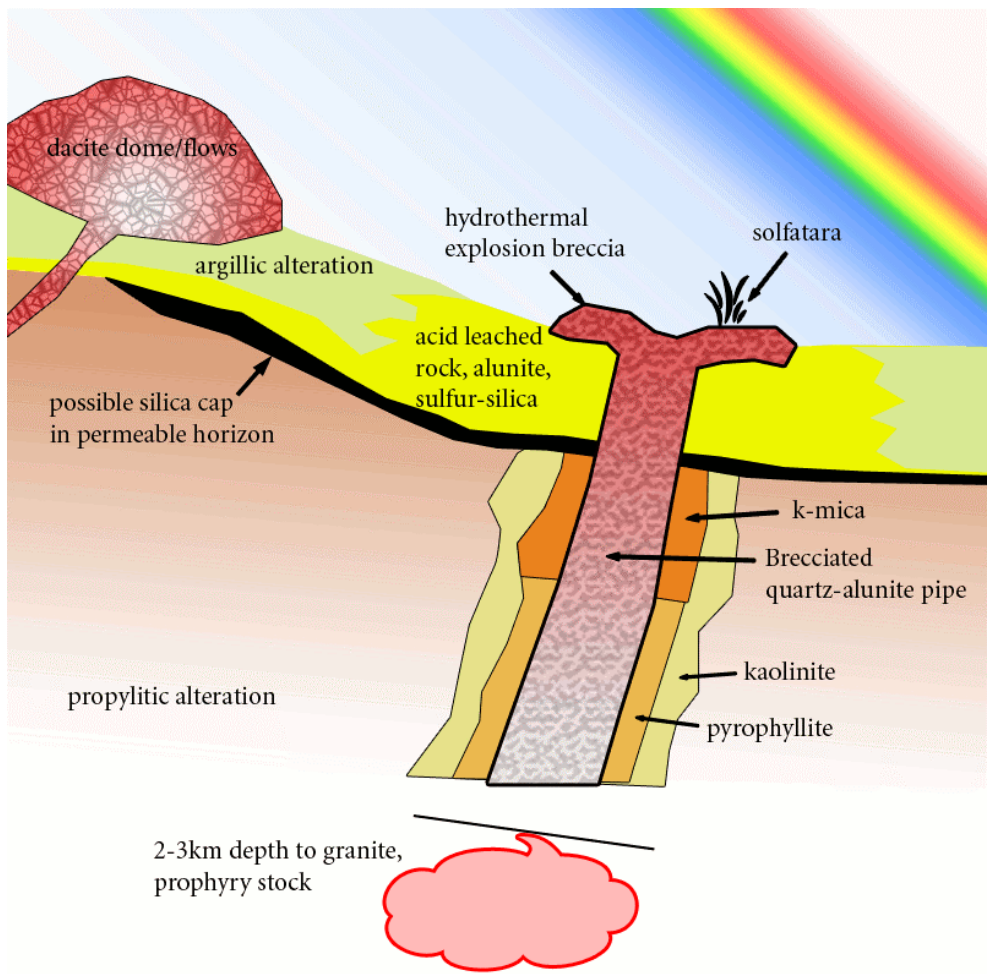

Figure 8 – Schematic cross section of epithermal high sulfur type deposit associated with granodiorite magmatism. After (Bonham Jr. and Giles, 1983).

2. <u>Erosional Phase</u>. Following the deposition of the Panorama Formation and cooling of the magma source, an erosional phase began. It is probable that the upper parts of the Panorama Formation volcanics were lost to erosion, and thin subaerial deposits were then emplaced upon an unconformity, followed by the Kelly Group rocks of the SPC.

3. <u>Depositional Phase</u>. The rocks of the Kelly Group were laid down all over the East Pilbara on top of a regional unconformity, trapping and preserving evidence of the Panorama hydrothermal event and thin subaerial erosional units below. The lower parts of the Kelly Group, including the SPC, were extensively silicified, perhaps in low temperature alteration caused by the later eruption of the ultramafic/mafic Euro Basalt.

We have proposed a genetic model for the Panorama Formation and surrounding units within the study region, which is summarized schematically in Figure 9.

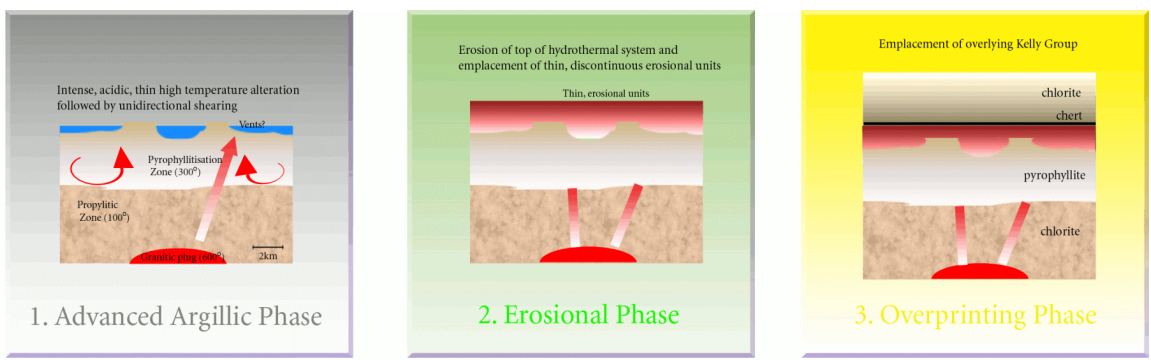

Figure 9 – Genetic model for the study region. See text for discussion.

## CONCLUSION

The hyperspectral maps presented here show that pyrophyllite is not limited to the paleosurface, but extends continuously down deep (several km) subvertical veins where muscovite eventually becomes prevalent. This suggests a syngenetic nature between subvertical and paleosurface (or shallow subsurface) pyrophyllite alteration. The remote sensing data favor a model where the veins fed hydrothermal fluids responsible for the pyrophyllite alteration in the horizontal paleosurface and vertical veins.

Acid-sulfate hydrothermal systems typically produce sulfates such as alunite and jarosite in their surface expressions (White and Hedenquist, 1990). This study has not revealed the location of any associated sulfate minerals, although their presence (even in small abundance) would do much to support the hydrothermal model presented here. If they have not been totally eroded, they would be ideal targets for further exploration in the region of the Panorama Formation.

## ACKNOWLEDGEMENTS

We thank the Geological Survey of Western Australia, particularly Tim Griffin, Brian Moore, Arthur Hickman, Martin Van Kranendonk and Kath Grey, for their generous assistance without which the fieldwork phase of this project would not have been possible. We also thank the GEMOC Analysis Unit, particularly Norm Pearson, Carol Lawson and Tin Tin Win, at Macquarie University for assistance with XRD and EMP measurements.